\documentclass[lettersize,journal]{IEEEtran}
\usepackage{amsmath,amsfonts}
\usepackage{algorithmic}
\usepackage{algorithm}
\usepackage{array}
\usepackage{textcomp}
\usepackage{stfloats}
\usepackage{url}
\usepackage{verbatim}
\usepackage{subcaption}
\usepackage{graphicx}
\usepackage{cite}
\usepackage{booktabs}
\usepackage{hyperref}
\usepackage{multirow}

\usepackage[dvipsnames]{xcolor}

\colorlet{blue}{black}

\usepackage{etoolbox}
\makeatletter
\patchcmd{\@makecaption}
  {\scshape}
  {}
  {}
  {}
\makeatother

\begin{document}

\title{DeSTA2.5-Audio: Toward General-Purpose Large Audio Language Model with Self-Generated Cross-Modal Alignment}


\author{\normalsize	
Ke-Han Lu, Zhehuai Chen, Szu-Wei Fu, Chao-Han Huck Yang, Sung-Feng Huang,
Chih-Kai Yang, Chee-En Yu, 

Chun-Wei Chen, Wei-Chih Chen, Chien-yu Huang,
Yi-Cheng Lin, Yu-Xiang Lin, Chi-An Fu, Chun-Yi Kuan, Wenze Ren, Xuanjun Chen, Wei-Ping Huang,
En-Pei Hu, Tzu-Quan Lin, Yuan-Kuei Wu, Kuan-Po Huang, Hsiao-Ying Huang, Huang-Cheng Chou, Kai-Wei Chang, Cheng-Han Chiang, 
Boris Ginsburg, Yu-Chiang Frank Wang, Hung-yi Lee 
}



\maketitle

\begin{abstract}
We introduce DeSTA2.5-Audio, a general-purpose Large Audio Language Model (LALM) designed for robust auditory perception and instruction-following.
Recent LALMs augment Large Language Models (LLMs) with auditory capabilities by training on large-scale audio-instruction datasets. However, existing LALMs have often suffered from the catastrophic forgetting of the LLM's original abilities. Therefore, balancing knowledge retention and audio perception has become a critical challenge.
To address this, we revisit the data construction pipeline and propose a self-generated cross-modal alignment strategy in which the backbone LLM generates its own training targets, named DeSTA. This approach aims at preserving the LLM’s native language proficiency thereby enabling zero-shot generalization without task-specific tuning.
We construct DeSTA-AQA5M, a large-scale, task-agnostic dataset containing 5 million training samples derived from 7,000 hours of audio spanning 50 diverse datasets, including speech, environmental sounds, and music. DeSTA2.5-Audio achieves state-of-the-art or competitive performance across a wide range of audio-language benchmarks, including Dynamic-SUPERB, MMAU, SAKURA, Speech-IFEval, and VoiceBench.
Comprehensive comparative studies demonstrate that our self-generated strategy outperforms existing training strategies.
Our findings underscore the importance of carefully designed data construction in LALM development and offer practical insights for building robust, general-purpose LALMs.
\footnote{https://github.com/kehanlu/DeSTA2.5-Audio}

\end{abstract}

\section{Introduction}

\begin{figure}[t]
    \centering
    \includegraphics[width=0.93\linewidth]{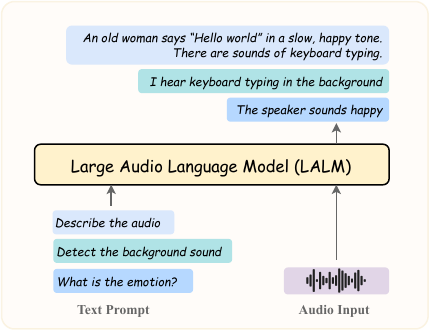}
    \caption{A general-purpose Large Audio Language Model (LALM) understands multifaceted audio information and accepts arbitrary text prompts to generate corresponding responses, enabling seamless multimodal interaction.}
    \vspace{-1em}
    \label{fig:LALM}
\end{figure}

\IEEEPARstart{T}{he}  development of general-purpose artificial intelligence has become a central focus in contemporary AI research, driven by the remarkable performance of large language models (LLMs) across various natural language understanding and generation tasks~\cite{achiam2023gpt, team2023gemini, touvron2023llama, dubey2024llama, bai2023qwen, yang2024qwen25,team2025gemma}.
Building on these advancements, a promising direction is to equip LLMs with multi-modal understanding capabilities, leading to the emergence of Large Audio Language Models (LALMs)~\cite{ lu24c_interspeech, desta2, gong2023joint, gonglisten, tang2024salmonn, ghosh-etal-2024-gama, hu2024wavllm, pmlr-v235-kong24a, chu2023qwen,chu2024qwen2, abouelenin2025phi, huang2025speechcaps, ghosh2025audio,held2024distilling, kuan2024speech} and Large Vision Language Models (LVLMs)\cite{lin2024vila, pmlr-v202-li23q, 11007678, zhuminigpt, gao2023llama}.
This paper focuses on building a general-purpose LALM, illustrated in Figure~\ref{fig:LALM}.

To develop a general-purpose LALM, two core capabilities are essential: \textbf{auditory perception} and \textbf{instruction-following}. Auditory perception refers to the comprehensive processing of auditory information, including speech, non-verbal cues, background sounds, and music. In addition, instruction-following entails interpreting diverse user commands, integrating them with internal knowledge, and producing contextually appropriate responses.
To achieve this, most LALMs adopt a modular architecture, combining a pre-trained audio model with a text-based LLM as the knowledge backbone~\cite{ lu24c_interspeech, desta2, gong2023joint, gonglisten, tang2024salmonn, ghosh-etal-2024-gama, hu2024wavllm, pmlr-v235-kong24a, chu2023qwen,chu2024qwen2, abouelenin2025phi, huang2025speechcaps, ghosh2025audio,held2024distilling}. These components are integrated through a cross-modal alignment process, which involves joint fine-tuning on a corpus of paired audio and text instructions to bridge the representational gap between the audio and text modalities. 
Accordingly, the research problem centers on training the fusion model with limited audio-related data, while leveraging the LLM’s generalization capabilities to handle unseen tasks.

{\color{blue}
However, acquiring high-quality data for cross-modal alignment remains a significant challenge. A common strategy involves transforming existing audio datasets by converting them into audio instruction-tuning datasets, consisting of (1) audio input, (2) textual instructions, and (3) textual responses. These training triplets can be generated either by text-based LLMs or through manual annotation.
While current LALMs perform well across a range of audio-related benchmarks~\cite{sakshi2025mmau, yang2025sakuramultihopreasoninglarge, chen2024voicebench, wang2024audiobench, 10448257, huang2025dynamicsuperb, yang2025towards}, many studies highlight the issue of catastrophic forgetting~\cite{lu24c_interspeech,desta2,tang2024salmonn,gonglisten,gong2023joint,hu2024wavllm, Goodfellow2024forgetting, lu2025speechifeval}, where the models become overly specialized to a limited set of audio-related tasks at the cost of generalization to unseen tasks.
Moreover, our previous work, Speech-IFEval~\cite{lu2025speechifeval}, also reveals that most LALMs experience significant degradation in text-based knowledge. 
Therefore, balancing text-based knowledge retention and audio generalization becomes a key challenge.}

{
\color{blue}
To mitigate catastrophic forgetting in LALM development, we revisit the paradigm of data construction process. We observe that a primary cause of degradation may stem from heterogeneous training targets. Specifically, previous studies often rely on training pairs generated by powerful LLMs~\cite{gonglisten, radhakrishnan2023whispering,gong2023joint, chu2023qwen,chu2024qwen2,ghosh-etal-2024-gama, chenaudio} or human-designed pairs~\cite{tang2024salmonn, hu2024wavllm}. 
While straightforward, we observe that these approaches inevitably introduce a distribution shift regarding the backbone LLM's native behavior, such as distinct stylistic and behavioral patterns. 
Consequently, training LALM on such heterogeneous data compels the model to simultaneously learn auditory perception and stylistic patterns from the foreign annotator. This implicit adaptation interferes with the cross-modal alignment objective, diverting the model’s focus and gradually overwriting its inherent text knowledge.
}

\begin{figure*}[t]
    \centering
    \includegraphics[width=0.98\linewidth]{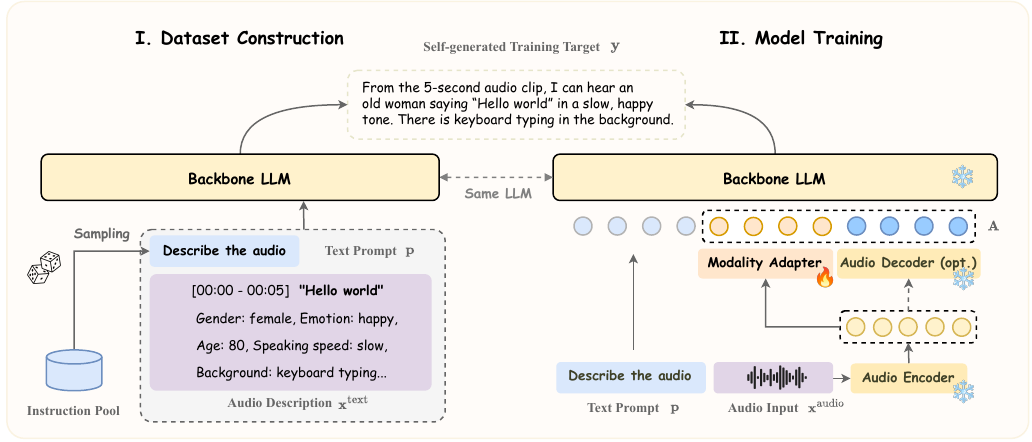}
    \caption{\textit{(Left)} \textbf{Dataset Construction:} An audio description $x^\text{text}$ and a randomly sampled prompt $\mathbf{p}$ are fed into the backbone LLM to generate training targets $\mathbf{y}$. \textit{(Right)} \textbf{Model Training:} The fusion model is trained using the self-generated targets $\mathbf{y}$ along with the corresponding audio inputs $x^\text{audio}$ and prompt $\mathbf{p}$. The fire and snowflake icons indicate trainable and frozen modules, respectively. The audio decoder is optional.}
   \label{fig:LALM-method}
\end{figure*}

Building on this insight, we introduce DeSTA\footnote{DeSTA stands for \textbf{De}scriptive \textbf{S}peech-\textbf{T}ext \textbf{A}lignment, first introduced in \cite{lu24c_interspeech} and further improved with self-generated design in \cite{desta2}. While the method generalizes beyond speech in this study, we retain the original naming convention for consistency.}, a self-generated cross-modal alignment framework that mitigates conflicting supervision by having the backbone language model generate its own training targets. 
Specifically, we convert the metadata of each audio segment into a structured textual description and pair it with an arbitrary prompt. Then, the LLM generates a corresponding response, which serves as the training target for cross-modal alignment. This self-generated supervision ensures stylistic and semantic consistency with the LLM’s native output distribution, thereby preserving its instruction-following capabilities.
{\color{blue}In our prior work, DeSTA2~\cite{desta2}, we demonstrate that training solely on a single general task, speech captioning, is sufficient for learning robust audio grounding. The resulting model exhibited strong zero-shot generalization across a wide range of downstream tasks without any task-specific instruction tuning.
In this work, we extend DeSTA by scaling our dataset construction pipeline across diverse audio domains, including speech, environmental sounds, and music. We aggregate 50 publicly available audio datasets spanning a wide range of audio information to create a large-scale corpus of approximately 5 million audio-text pairs, which we call DeSTA-AQA5M. This dataset encompasses various forms of audio information, including paralinguistic cues, speaker identities, audio quality indicators, and environmental or contextual sounds.
This expanded model, referred to as DeSTA2.5-Audio, further validates the effectiveness of self-generated alignment in enabling general and robust audio-language modeling.}

{\color{blue}
We evaluate DeSTA2.5-Audio on several audio-language benchmarks, including Dynamic-SUPERB~\cite{huang2025dynamicsuperb, 10448257}, MMAU~\cite{sakshi2025mmau}, SAKURA~\cite{yang2025sakuramultihopreasoninglarge}, Speech-IFEval~\cite{lu2025speechifeval}, and VoiceBench~\cite{chen2024voicebench}. We also conduct CLAP-style analysis to study the learned ``modality-gap'' at the representation-level~\cite{10095889, 10095969}. Despite using only 7,000 hours of training data, compared to 510,000 hours used by the baseline Qwen2-Audio-Instruct~\cite{chu2024qwen2}, our model achieves state-of-the-art or competitive results across domains.
}
Our comparative analysis also emphasizes the critical role of the distribution mismatch problem. Models trained on self-generated targets consistently outperform those trained on mismatched data sources (e.g., from other LLM), even when the latter originate from larger or more powerful LLMs. Although adaptation techniques like LoRA~\cite{hulora} can alleviate distribution mismatches, they often come at the cost of reduced performance in complex reasoning or out-of-domain scenarios. These results suggest that training data distribution should be a fundamental design consideration in the development of LALMs.

The contribution can be summarized as follows:

\begin{itemize}
\color{blue}
\item We propose DeSTA, a self-generated cross-modal alignment method that reduces the mismatch between training data and the model’s original behavior. We prevent the catastrophic forgetting problem and eliminate the need for task-specific instruction-tuning.

\item We introduce a large-scale audio instruction-tuning dataset, DeSTA-AQA5M, which covers a wide array of audio features and has 5 million audio-prompt-response triplets. The dataset comprises 7,000 hours of diverse audio data spanning speech, environmental sounds, and music from 50 publicly available audio datasets.

\item Our model, DeSTA2.5-Audio, achieved superior results across multiple audio-language benchmarks. We demonstrate strong performance on both auditory perception and knowledge-intensive tasks, showing significantly enhanced robustness compared to existing LALMs.

\item We conducted extensive comparison studies validating that self-generated design is critical for success, revealing significant limitations in previous LALMs. Our findings reshape the design philosophy and provide important insights for future research.

\end{itemize}

\IEEEpubidadjcol

\section{Related works}

\subsection{Large Audio Language Model (LALM)}

There are already several overview papers about LALMs; Therefore, we briefly review only the technologies directly relevant to this work. For a comprehensive overview, please refer to these survey papers~\cite{arora2025landscapespokenlanguagemodels,ji2024wavchatsurveyspokendialogue,peng2025surveyspeechlargelanguage,cui2025recentadvancesspeechlanguage,yang2025towards}.


{
\color{blue}
Recent research on general-purpose LALMs has focused on addressing the text-audio modality gap by fine-tuning fusion modules with audio–text pairs~\cite{desta2, lu24c_interspeech, gong2023joint, gonglisten, tang2024salmonn, ghosh-etal-2024-gama, hu2024wavllm, pmlr-v235-kong24a, chu2023qwen,chu2024qwen2, abouelenin2025phi, huang2025speechcaps, ghosh2025audio, held2024distilling}. 
Therefore, the availability and construction of training datasets play a critical role in model development. 
To efficiently create audio-text pairs, synthetic data generated by powerful text-based LLMs is widely adopted. This approach involves repurposing existing audio datasets and utilizing their annotated textual metadata to construct input-output pairs.
For instance, LTU~\cite{gonglisten} prompts LLMs to generate open-ended audio question-answer pairs. Qwen-Audio~\cite{chu2023qwen} creates interactive data to enhance chat behavior. DeSTA~\cite{lu24c_interspeech} augments metadata into speech captions as a pre-training stage, enabling the model to learn multifaceted speech descriptions before the instruction-tuning phase.
These synthetic processes can be easily automated and scaled. Once the initial textual metadata is collected, the training data can be generated accordingly.} Consequently, scaling up audio datasets has emerged as another key area of research~\cite{xue2025audio, pandey2025sift, mei2023wavcaps, he2024emilia}.
Beyond audio-to-text models, an emerging research direction explores multi-modal LLMs designed for interactive scenarios that require both audio understanding and speech generation~\cite{fu2025vita,defossez2024moshi,zeng2024glm,achiam2023gpt,yang2024building}. These models often build upon the architectural and training insights from LALMs, while placing greater emphasis on generating natural and context-aware speech responses.

\subsection{Self-Generation Mechanism}

The concept of self-generated data has gained significant traction in text LLM fine-tuning~\cite{wang-etal-2023-self-instruct, honovich-etal-2023-unnatural, liself}. 
Rather than relying on supervision from a powerful teacher model or human-annotated data, self-generated approaches enable the model to create its own training data. 
A key advantage of this method is that the training targets are naturally matched with the model’s inherent behavior and data distribution. 
Several studies on text LLM have shown that this alignment can improve performance~\cite{ren-etal-2024-learn, li-etal-2024-quantity,yang2024self,chiang2025tract,critiqueoutloud}.

In multi-modal contexts, researchers have begun to adapt this approach to the speech domain. 
One common method uses automatic speech recognition (ASR) transcriptions as inputs and allows the backbone LLM to generate textual continuations as training targets~\cite{fathullah2023audiochatllama, wang2023blsp, yu-etal-2024-self-powered, hsiao2025analyzing}. 
However, these efforts are often limited to spoken-content-centric tasks, such as ASR and spoken question answering, and tend to underutilize the rich, multidimensional information present in audio. 
{\color{blue}
Our prior work, DeSTA2~\cite{desta2}, extended the self-generated paradigm in the speech domain by incorporating paralinguistic cues, speaker characteristics, and other non-verbal audio signals. 
We found that training on self-generated “speech captions,” which provide comprehensive audio grounding, enables the model to yield strong zero-shot performance on a variety of tasks, demonstrating the ability to generalize from a single training objective.
Moreover, these studies rarely analyze how model behavior or performance might differ if the training data were not self-generated. 
In this work, we further broaden our approach to encompass a wider variety of audio information, and we provide comprehensive evaluations and analyses to validate its effectiveness.
}

{ \color{blue} In summary, while using text metadata to synthesize audio-text training pairs has been widely adopted in existing LALM studies, our approach differs fundamentally in design philosophy. Compared to existing LALMs such as LTU~\cite{gonglisten}, SALMONN~\cite{tang2024salmonn} and Qwen-Audio~\cite{chu2023qwen,chu2024qwen2} that rely on instruction-tuning datasets generated by heterogeneous sources, our design prioritizes a more streamlined and robust alignment strategy. We not only simplify the training pipeline, but also fundamentally avoid the trade-off between auditory perception and textual knowledge retention. }

\begin{table*}[!t]
\footnotesize
\renewcommand{\arraystretch}{1.3}
\setlength{\tabcolsep}{3pt}
\color{blue}

\caption{
\color{blue}
{Data examples from the dataset construction stage. We feed the audio description and prompt to generate training targets. These examples also show that identical inputs lead to different model responses across LLMs, highlighting the distribution mismatch problem we explored in this paper.}}
\small	
\label{tab:data_examples}
\centering
\begin{tabular}{l}
\toprule


\textbf{Audio Description:} {[00:00-00:03] If not me, then who?} \\{(Emotion: Sad, Gender: Female, Pitch: Normal, Volume: Very low, Speaking speed: Slow, Duration: 3s)}\\
\textbf{Prompt:} Describe the core of the clip with brevity. \\

\textbf{Model Response (Training Target)} \\
\quad $\bullet$ \textbf{Llama3.1:} A melancholic, introspective tone, questioning one's identity and purpose. \\
\quad $\bullet$ \textbf{Qwen2.5:} A sad female voice asks, "If not me, then who?" in a very low, slow, and normal-pitched tone. \\
\quad $\bullet$ \textbf{Gemma3:} A woman expresses a feeling of sadness and questioning her worth. \\
\midrule
\textbf{Audio Description:} [00:00-00:10] (A helicopter vibrates loudly.) \\
\textbf{Prompt:} Describe this audio's character. \\
\textbf{Model Response (Training Target):}\\
\quad $\bullet$ \textbf{Llama3.1:} The audio clip is a loud, vibrating sound, likely caused by the rotor blades of a helicopter in operation. \\
\quad $\bullet$ \textbf{Qwen2.5:} The audio features the character of a loud, vibrating helicopter. \\
\quad $\bullet$ \textbf{Gemma3:} Loud, vibrating, and sounds like a helicopter. \\

\bottomrule
\end{tabular}
\end{table*}

\section{Methodology}

An overview of the DeSTA framework is shown in Figure~\ref{fig:LALM-method}.
Just as humans can imagine auditory scenes from written narratives, such as voices and ambient sounds in a novel or film script, modern LLMs exhibit similar abilities. Most auditory concepts, such as tone, speaking style, and environmental sounds, can be described in natural language. Based on this intuition, DeSTA aims to provide an approximate textual description of audio input, which will be used to generate the corresponding training target using the backbone LLM.

\subsection{Self-Generated Dataset Construction}
\label{sec:dataset_construction}
{\color{blue}During the data construction phase, we first collect a diverse set of datasets containing detailed metadata. 
As illustrated in Table~\ref{tab:data_examples}, we empirically convert the metadata into a structured textual format following the schema below: 
\begin{quote}
\texttt{[timestamp] Spoken content (non-verbal attribute name: value)}
\end{quote} 
\textcolor{blue}{The resulting initial dataset, denoted as $\mathcal{D}_\text{initial} = \{({x}^\text{audio}, {x}^\text{text})\}$, pairs each audio clip ${x}^\text{audio}$ with its corresponding textual description ${x}^\text{text}$.}

Based on these initial audio–description pairs, we utilize the backbone LLM to generate training targets that reflect how the model would naturally respond to the given descriptions.
Specifically, we randomly sample a text prompt $\mathbf{p}$ from a predefined instruction pool $\mathcal{P}$, which includes diverse prompt types such as descriptive tasks (e.g., ``Describe the audio"), role-playing scenarios (e.g., ``Respond to the audio based on its expression"), and open-ended questions (e.g., ``Where is the audio being recorded?"). These prompts are crafted to maximize the LLM’s ability to utilize all available information from the textual description, thereby enhancing the quality of cross-modal alignment.
The backbone LLM then takes the text description $x^\text{text}$ and the prompt $\mathbf{p}$ as inputs and produces a response $\mathbf{y} = \text{LLM}(x^\text{text}, \mathbf{p})$. This automated setup enables the generation of detailed and context-aware responses, eliminating the need for labor-intensive, task-specific instruction design.
Through this procedure, we construct an audio instruction-tuning dataset in the form of $\mathcal{D} = {(x^\text{audio}, x^\text{text}, \mathbf{p}, \mathbf{y})}$, where each data point includes the audio input, its corresponding text description, the sampled prompt, and the LLM-generated response.
}

\subsection{Model Training}

As illustrated in Figure~\ref{fig:LALM-method} (Right), our model adopts a modular architecture that integrates a pre-trained audio model with an instruction-tuned LLM. To bridge the audio and language modalities, we insert a modality adapter composed of Q-Former blocks between the two modules. In our design, we freeze all parameters of the audio model and the LLM, and fine-tune only the modality adapter to learn robust audio–text alignment representations. The fusion model is fine-tuned on triplets in the form $(x^{\text{audio}}, \mathbf{p}, \mathbf{y})$.

The audio input $x^{\text{audio}}$ is first encoded into a continuous representation by applying Q-Former blocks~\cite{pmlr-v202-li23q} to multiple intermediate hidden states from the audio encoder~\cite{lu24c_interspeech, desta2}. Formally, let $\mathbf{h}^{(\ell)} \in \mathbb{R}^{T \times d}$ denotes the output from the $\ell$-th encoder layer, where $T$ represents the temporal dimension and $d$ is the hidden dimension. For each selected layer $\ell$, we define learnable query vectors $\mathbf{Q}^{(\ell)} \in \mathbb{R}^{N \times d}$, where $N$ is the number of queries. The Q-Former computes:
\[
\mathbf{f}^{(\ell)} = \text{Q-Former}(\mathbf{Q}^{(\ell)}, \mathbf{h}^{(\ell)}) \in \mathbb{R}^{N \times d}
\]

The outputs from multiple layers are aggregated using learnable scalar weights $\alpha^{(\ell)}$ with the constraint $\sum_\ell \alpha^{(\ell)} = 1$~\cite{10502279}, and subsequently projected through a linear layer to match the LLM embedding dimension $d'$:
\[
\mathbf{F} = \text{Linear}\left(\sum_{\ell} \alpha^{(\ell)} \mathbf{f}^{(\ell)}\right) \in \mathbb{R}^{N \times d'}
\]
Optionally, a linguistic representation may be incorporated by transcribing the input audio $x^{\text{audio}}$ using the audio decoder to obtain a text sequence $\mathbf{t} \in \mathbb{R}^L$, where $L$ denotes the sequence length. This transcription is then passed through the LLM's token embedding layer:
\[
\mathbf{E} = \text{Embed}(\mathbf{t}) \in \mathbb{R}^{L \times d'}
\]
When used, the discrete features $\mathbf{E}$ are concatenated with the continuous features $\mathbf{F}$ along the sequence dimension to form the final audio representation:
\[
\mathbf{A} = [\mathbf{F}; \mathbf{E}] \in \mathbb{R}^{(N + L) \times d'}
\]
Otherwise, only the continuous features $\mathbf{F}$ are used.

The resulting audio embeddings $\mathbf{A}$ are passed to the LLM along with the prompt embeddings $\mathbf{P}$ to autoregressively generate the output sequence:
\[
y_i = \text{LLM}(\mathbf{P}, \mathbf{A}, y_{<i})
\]
where $y_i$ represents the $i$-th token in the generated sequence, and $y_{<i}$ denotes all previously generated tokens. The model undergoes end-to-end optimization using the standard next-token prediction loss computed on the training targets $\mathbf{y}$.

{
\color{blue}
The key advantage of the self-generated design in DeSTA is that it directs the optimization process to focus exclusively on cross-modal alignment.
As an illustration, the optimization process for the trainable parameters can be conceptualized in two objectives: 
\begin{itemize}
    \item \textbf{Cross-Modal Grounding}: maps audio features to the LLM's semantic space
    \item \textbf{Stylistic Adaptation}: shifts the output distribution to match the target's specific phrasing and style
\end{itemize}
When training targets are derived from external sources (e.g., human annotators or different LLMs), they inevitably diverge from the backbone LLM's native distribution, introducing a divergent training signal that gradually distorts the model behavior. In contrast, by employing the backbone LLM itself to generate training targets, we eliminate this stylistic mismatch. This interpretation aligns with our empirical findings and provides a rationale for the robust generalization and instruction-following abilities observed in DeSTA2.5-Audio.
}

\section{Experiment Setup}

\subsection{Dataset}
To construct the training corpus, we collected 50 publicly available datasets spanning a wide range of domains in audio processing. We prioritize datasets containing metadata that cover a comprehensive range of audio information, 
including paralinguistic features (\textit{e.g.,} pitch, loudness, speaking rate, prosody, timbre, emotional tone, and speaking style), speaker identity attributes (\textit{e.g.,} accent, gender, and age), audio quality indicators (\textit{e.g.,} background noise levels, reverberation, and spoofed or synthetic audio), and environmental or contextual sounds (\textit{e.g.,} animal vocalizations, human actions, ambient sounds, musical instruments, music genres, and natural environments). While the majority of the data are in English, some datasets include Mandarin or code-switching audio samples.

In total, the dataset comprises approximately 7,000 hours of audio: 5,400 hours of speech, 1,000 hours of environmental sounds, and 500 hours of music.
For a detailed breakdown of the metadata used in each dataset, we refer readers to our project page for details\footnotemark[1]. The datasets used in this study are categorized as follows:

\paragraph{\textbf{Speech domain}} IEMOCAP~\cite{Busso2008}, DailyTalk~\cite{10095751}, GLOBE~\cite{wang2024globe}, VCTK-corpus~\cite{yamagishi2019cstr}, MELD~\cite{poria-etal-2019-meld}, PromptTTS~\cite{10096285}, Expresso~\cite{nguyen23_interspeech}, AccentDB~\cite{ahamad-anand-bhargava:2020:LREC}, VoxCeleb1~\cite{Nagrani17}, Anispeech~\cite{shoukanlabs_anispeech_2024}, MUSAN~\cite{snyder2015musan}, MSP-IMPROV~\cite{Busso_2017}, Fair-speech~\cite{veliche24_interspeech}, CREMA-D~\cite{6849440}, CAFE~\cite{10.1145/3204949.3208121}, EMOVO~\cite{costantini2014emovo}, Speech accent archive~\cite{weinberger2011speech}, EMNS~\cite{EMNS_corpus}, KeSpeech~\cite{tang2021kespeech}, ESD~\cite{zhou2021seen}, LibriSpeech-c~\cite{lin-etal-2024-continual}, L2Arctic~\cite{zhao18b_interspeech}, CommonVoice (English and Chinese)\cite{commonvoice:2020}, EmoV-DB~\cite{adigwe2018emotionalvoicesdatabasecontrolling}, LibriTTS-R~\cite{52510}, Dusha~\cite{kondratenko2022largerawemotionaldataset}, MSP-PODCAST~\cite{8003425}, AliMeeting~\cite{Yu2022M2MeT}, CSZS~\cite{10446737}, NTUML2021~\cite{10626762}, Speech Command~\cite{warden2018speech}, Libricount~\cite{DBLP:data/10/StoterCHE18}, Voxlingual~\cite{turian2022hear}, ASVspoof~\cite{9358099}, BIIC-Podcast~\cite{10388175}, and CodecFake ~\cite{wu24p_interspeech, chen2025codecfake}.
\paragraph{\textbf{Environmental sound domain}} Audioset~\cite{7952261}, AudioCaps~\cite{kim-etal-2019-audiocaps}, Wavcaps~\cite{mei2023wavcaps}, Clotho~\cite{drossos2020clotho}, VocalSound~\cite{gong_vocalsound}, UrbanSound8k~\cite{Salamon:UrbanSound:ACMMM:14}, ESC50~\cite{piczak2015dataset}, FSD50K~\cite{fonseca2021fsd50k} , and THMINT-QI~\cite{zezario2023study}.
\paragraph{\textbf{Music domain}} Nsynth~\cite{engel2017neural}, OpenSinger~\cite{huang2021multi}, FMA~\cite{defferrard2017fma}, GTZAN~\cite{tzanetakis2002musical} , and Mridangam~\cite{anantapadmanabhan2013modal}.

Regarding the instruction pool, we curated 4,000 prompts for the speech category and 3,000 prompts for the environmental sound and music categories. Due to the heterogeneity and varying sizes of the source datasets, we applied an upsampling strategy to balance across domains. Each audio sample is paired with multiple prompts, formatted as:
$$
\{x^{text}\} \{p\}
$$
All responses were generated using the vLLM toolkit~\cite{kwon2023efficient}, with decoding parameters set to a temperature of 0.05 and a top-p value of 1.0. This process yielded a large-scale dataset of approximately 5 million audio–prompt–response triplets, which we refer to as DeSTA-AQA5M. This dataset serves as the training corpus for DeSTA2.5-Audio.
{\color{blue}
In this work, while we primarily study the performance of Llama3.1-8B-Instruct~\cite{dubey2024llama}, the same data construction process can be easily generated by other LLMs within one day on a single NVIDIA A100-80GB GPU.
}

\subsection{Model Specification and Training Setup}
We adopt Llama3.1-8B-Instruct and Whisper-large-v3 as the foundational components. We employ a six-layer Q-former architecture~\cite{pmlr-v202-li23q} with 64 queries as the modality adapter.
The query vectors attend to intermediate hidden states from Whisper encoder layers 8, 16, 24, and 32, allowing the model to capture multi-scale acoustic features~\cite{yang21c_interspeech,10502279}.
For the optional linguistic representation, we provide offline transcriptions for datasets originating from the speech domain. For datasets in the audio and music domains, no transcription is used, and only continuous embeddings derived from the Q-Former are utilized during training. During inference, we employ a lightweight pre-trained voice activity detection (VAD) model~\cite{SileroVAD} to identify the presence of human speech in audio inputs and conditionally activate the Whisper decoder when necessary. 

The model implementation is based on the Transformers library~\cite{wolf-etal-2020-transformers}. 
It consists of 8.8 billion total parameters, with 131 million trainable parameters.
Training is performed for five epochs using the Adam optimizer~\cite{kingma2014adam}, a cosine annealing learning rate schedule, and 2000 warm-up steps. 
{\color{blue}The training is conducted on a cluster of 8 NVIDIA A100-80GB GPUs for five days, with a global batch size of 96 and an initial learning rate of 1e-4. The total number of training steps is approximately 250,000.
}

\subsection{Comparison Study}

To assess the robustness and effectiveness of our proposed framework, we construct a controlled subset of 500,000 samples from DeSTA-AQA5M. This subset enables systematic analysis of how various training target generation alternatives and model configurations affect model performance. 
All comparison experiments are trained for ten epochs using identical model architectures and hyperparameter settings.

In the dataset-level comparison, we fix both the audio descriptions and prompts, and vary only the LLM responsible for generating the training targets. We consider Llama3.1-8B-Instruct~\cite{dubey2024llama}, Qwen2.5-7B-Instruct~\cite{yang2024qwen25}, Gemma3-12B-it~\cite{team2025gemma}, and Llama3.1-70B-Instruct~\cite{dubey2024llama}. This setup facilitates an analysis of how variations in the output distribution across different LLMs influence downstream training outcomes under otherwise controlled conditions. Additionally, to isolate the impact of prompt diversity, we construct a caption-style dataset using a single fixed prompt (i.e., “What can you hear from the audio?”) as proposed in our prior work~\cite{desta2}.
{
\color{blue}
For the model-level comparison, we augment the LLM backbone with LoRA adapters~\cite{hulora} of rank 8, applied to the query, key, and value projections of each transformer layer. This increases the number of trainable parameters to 140 million.
}


\section{Evaluation Setup}

\subsection{Benchmarks}

To evaluate the capabilities of LALM across instruction-following, perceptual understanding, and reasoning, we adopt a diverse suite of benchmarks.
These benchmarks are designed to target distinct challenges in audio-language learning and, when combined, offer a comprehensive evaluation of model behavior across task types, domains, and reasoning levels.

\textbf{Dynamic-SUPERB Phase-1}~\cite{10448257} evaluates instruction-following and speech understanding across 48 classification tasks. These tasks are categorized into five groups: content (CON), semantic (SEM), paralinguistic (PAR), degradation (DEG), and speaker (SPK). Performance is measured by classification accuracy against ground truth labels.

\textbf{Dynamic-SUPERB Phase-2}~\cite{huang2025dynamicsuperb} extends the benchmark to 180 tasks by incorporating new contributions from the research community, including regression and open-ended generation tasks across the speech, environmental sound, and music domains. Performance is assessed using task-specific metrics, following the official evaluation guidelines.

{
\color{blue}
\textbf{MMAU}~\cite{sakshi2025mmau} is a benchmark for evaluating advanced audio-language understanding and reasoning across speech, environmental sounds, and music, using multiple-choice question formats. Some questions require expert-level domain knowledge for correct interpretation. We use the test-mini subset, and the performance is evaluated by accuracy against ground truth answers.
}

{
\color{blue}
\textbf{SAKURA}~\cite{yang2025sakuramultihopreasoninglarge} is a benchmark developed to evaluate single-hop and multi-hop reasoning in LALMs, referred to as SAKURA-Single and SAKURA-Multi, respectively. The benchmark consists of multiple-choice questions that include gender, emotion, animal and language classification tasks. Specifically, single-hop questions assess basic auditory perception, such as identifying sounds like ``What animal is making this sound?'' In contrast, multi-hop questions require combining auditory cues with world knowledge and reasoning beyond the immediate input, such as ``Is the sound source a mammal or not?''. These benchmarks are particularly useful for assessing the internal knowledge of LALM.
}

{
\color{blue}
\textbf{Speech-IFEval}~\cite{lu2025speechifeval} is a diagnostic benchmark designed to assess whether LALMs retain their instruction-following after cross-modal alignment. 
Unlike benchmarks that emphasize task-level accuracy, Speech-IFEval introduces the instruction-following rate (IFrate). This metric evaluates whether the model follows output constraints (e.g., limiting response length or generating valid JSON), independent of correct audio interpretation.
To quantify degradation in instruction-following, Speech-IFEval further defines the forgetting rate ($\Delta$) as the relative drop in IFrate between an LALM and its backbone LLM. A negative score indicates that the LALM has lost instruction-following ability compared to its text-only backbone.
}

\textbf{VoiceBench}~\cite{chen2024voicebench} comprises a collection of tasks designed to evaluate spoken interaction performance. It utilizes text-to-speech (TTS) systems to convert textual instructions into audio inputs, simulating realistic voice-based scenarios. Most tasks in VoiceBench are adapted from existing text-based benchmarks and are primarily content-oriented. We evaluate the performance in accordance with the original guidelines.

\subsection{Cascade Baselines}
We construct two cascade baselines that operate on text-only inputs. The first baseline, ASR+LLM, uses Whisper-large-v3 to transcribe the audio signal into text, which is then directly fed into Llama3.1-8B-Instruct. While this approach omits non-verbal acoustic information, ASR+LLM serves as a strong baseline across multiple benchmarks due to its complete preservation of the LLM’s pre-trained textual knowledge and reasoning ability.
To incorporate audio-based information beyond linguistic content, we introduce an extended variant that includes automatic audio captioning (AAC), denoted as ASR+AAC+LLM. Specifically, we use DeSTA2.5-Audio to generate a detailed caption of the audio, which is then concatenated with the ASR transcription to form a richer textual input.
This allows the model to reason more effectively about context, emotion, and acoustic scenes, potentially enhancing its performance on tasks that rely on paralinguistic and contextual cues.

\begin{table*}[!t]
\renewcommand{\arraystretch}{1.3}
\setlength{\tabcolsep}{3pt}

\caption{Evaluation results across Dynamic-SUPERB Phase-1, MMAU, SAKURA and Speech-IFEval. $\dagger$ signifies the cascade reference system result from \cite{lu2025speechifeval}, and $\uparrow$ denotes higher scores are better.}
\label{tab:main_results}
\centering
\begin{tabular}{lcccccccccccccc}
\toprule
\multirow{2}{*}{\textbf{Model}}
& \multicolumn{6}{c}{\textbf{Dynamic-SUPERB Phase-1 ($\uparrow$)}} 
& \multicolumn{4}{c}{\textbf{MMAU ($\uparrow$)}} 
& \multicolumn{2}{c}{\textbf{SAKURA ($\uparrow$)}}
& \multicolumn{2}{c}{\textbf{Speech-IFEval ($\uparrow$)}} \\
\cmidrule(lr){2-7}
\cmidrule(lr){8-11}
\cmidrule(lr){12-13}
\cmidrule(lr){14-15}
 & CON & SEM & PAR & DEG & SPK & \colorbox{lightgray}{Avg}
 & Speech & Sound & Music & \colorbox{lightgray}{Avg}
 & Single & Multi 
 & IFrate & $\Delta$ \\
\midrule
ASR+LLM
& 73.05 & 59.67 & 40.50 & 43.13 & 43.50 & 51.71 
& 57.96 & 38.44 & 46.41 & 47.60
& 40.20 & 37.40
& \multirow{2}{*}{$\text{93.52}^\dagger$} & \multirow{2}{*}{--}
\\

ASR+AAC+LLM
& 82.77 & 68.58 & 45.93 & 54.87 & 48.10 & 60.97
& 57.96 & 48.35 & 47.01 & 51.10
& 59.45 & 51.30
\\

\midrule

LTU-AS \cite{gong2023joint} & 43.95 & 36.00 & 17.14 & 37.53 &40.20 & 36.11
& 23.35 & 9.10 & 20.60 & 17.68
& 40.90 & 18.10
& 29.19 & -54.90
\\
SALMONN \cite{tang2024salmonn}
& 52.00 & 50.75 & 24.50 & 28.16 & 33.20 & 36.44
& 34.80 & 41.00 & 25.50 & 33.70
& 42.50 & 35.30
& 36.89 & -50.20
\\
Qwen-Audio \cite{chu2023qwen} & 61.77 & 47.17 & 28.64 & 30.95 & 41.40 & 40.79
& 30.03 & 44.00 & \textbf{55.25} & 43.10
& 73.20 & 46.90
& 32.98 & --
\\

Qwen2-Audio-Instruct \cite{chu2024qwen2} & 77.64 & 59.17 & 29.21 & 43.58 & 47.90 & 51.69
& 42.04 & 54.95 & 50.98 & 49.20
& \textbf{81.20} & 49.10
& 47.11 &  -- \\

DeSTA2 \cite{desta2}
& 79.41 & 59.42 & 43.14 & 51.63 & 42.50 & 56.78
& 55.86 & 47.75 & 47.60 & 50.40
& 63.00 & 57.10
& 89.23 & -3.57 \\

DeSTA2.5-Audio
& \textbf{88.95} & \textbf{66.00} & \textbf{63.71} & \textbf{65.08} & \textbf{56.10} & \textbf{69.53}
& \textbf{59.16} & \textbf{60.66} & 52.69 & \textbf{57.50}
& 76.65 & \textbf{69.85}
& \textbf{93.89} & \textbf{+0.40} \\

\bottomrule
\end{tabular}
\end{table*}

\subsection{Model Inference Setup}
%
While recent studies have demonstrated that careful prompt design, such as chain-of-thought prompting \cite{wei2022chain}, can significantly improve multi-modal performance \cite{ma2025audio, wu2023role}, our goal is to avoid confounding effects caused by prompt engineering. 
Therefore, we intentionally keep the setting simple throughout this study, and we leave the exploration of more advanced prompting techniques as a direction for future work. 

We utilize a set of standardized system prompts to guide the model's behavior across different evaluation scenarios. 
The default prompt is \textit{``Focus on the audio clips and instruction,"} which encourages the model to attend to both modalities and generate informative responses. For benchmarks involving multiple-choice formats, such as Dynamic-SUPERB Phase-1, MMAU, and SAKURA, we append the instruction \textit{``Choose one of the options without any explanation"} to ensure unambiguous predictions. In VoiceBench, where the instruction is spoken, we adopt the prompt \textit{``You are a voice assistant. Act as if you are a natural language partner"} to simulate more interactive behavior. For tasks evaluated with word-level metrics, such as ASR tasks using word error rate, we specify \textit{``Respond directly without any other words"} to suppress auxiliary phrases and ensure fair assessment. We use greedy decoding across all evaluations.


\begin{figure}[]
    \centering
    
    \begin{subfigure}{0.45\textwidth} 
        \centering
        \includegraphics[width=0.7\textwidth]{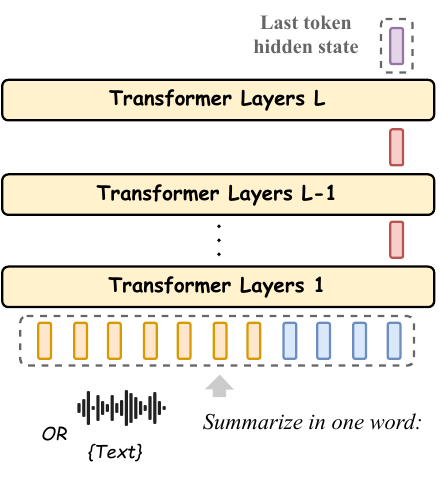} 
        \caption{Representation Extraction}
        \label{fig:representation:a}
    \end{subfigure}
    \hfill
    \begin{subfigure}{0.45\textwidth} 
        \centering
        \includegraphics[width=0.78\textwidth]{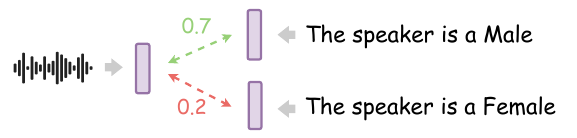} 
        \caption{Similarity-based Classification}
        \label{fig:representation:b}
        
    \end{subfigure}
    
    \caption{
    \color{blue}
    Illustration of the CLAP-style evaluation using the backbone LLM for embedding extraction. (a) \textbf{Representation Extraction}: Inputs are fed into the LLM. We utilize the hidden state of the last token as the aggregated representation for the input. (b) \textbf{Similarity-based Classification}: The derived audio representation is compared against the text representations of all candidate class labels using cosine similarity. The label with the highest similarity score (in this example, ``Male'') is selected as the prediction.}
    \label{fig:representation}
    \vspace{-1em}
\end{figure}

\subsection{Representation-Level Evaluation Setup}
\label{sec:analysis_study}

{
\color{blue}

Beyond standard instruction-based evaluation, we adopt a CLAP-style protocol using similarity-based classification~\cite{10095889, 10095969} to assess the representation-level cross-modal alignment quality, as illustrated in Figure~\ref{fig:representation}.
Since our autoregressive models do not yield a single explicit embedding vector for variable-length inputs, we utilize the backbone LLM as a universal encoder~\cite{jiang2024e5, hu25b_interspeech}. As depicted in Figure~\ref{fig:representation:a}, we feed either the audio (via the modality adapter) or the text sequence into the LLM. We then append a summarization prompt (e.g., ``summarize the audio in one word:'') to the input and extract the hidden state of the last token as the aggregated representation~\cite{jiang-etal-2024-scaling}.

We conduct this analysis on SAKURA-Single, targeting gender (2 classes), emotion (5 classes), and animal sounds (9 classes). Class labels are converted into text embeddings using descriptive templates (e.g., ``The speaker is a male'', ``The speaker feels happy'' and ``There is a cat''). As depicted in Figure~\ref{fig:representation:b}, classification is performed by identifying the text representation with the highest cosine similarity to the audio embedding. Unless otherwise specified, we report scores from the final transformer layer. Higher accuracy suggests that the LLM interprets the audio representations similar to the semantic concept.


}

\section{Main Results}

\subsection{Results on Dynamic-SUPERB Phase-1, MMAU, SAKURA and Speech-IFEval}
\label{sec:big_results}

Table~\ref{tab:main_results} presents results across benchmarks, compared to several representative models. 
To contextualize the benchmark performance, we first consider tasks related to auditory perception, including Dynamic-SUPERB Phase-1 and MMAU, which evaluate models on audio understanding. The SAKURA benchmark, which spans both single-hop and multi-hop reasoning, serves as a bridge between basic perception and higher-level reasoning. Finally, we assess general knowledge and instruction-following through benchmarks like SAKURA-Multi and Speech-IFEval.

\begin{figure*}[!t]
\centering
\includegraphics[width=.80\linewidth]{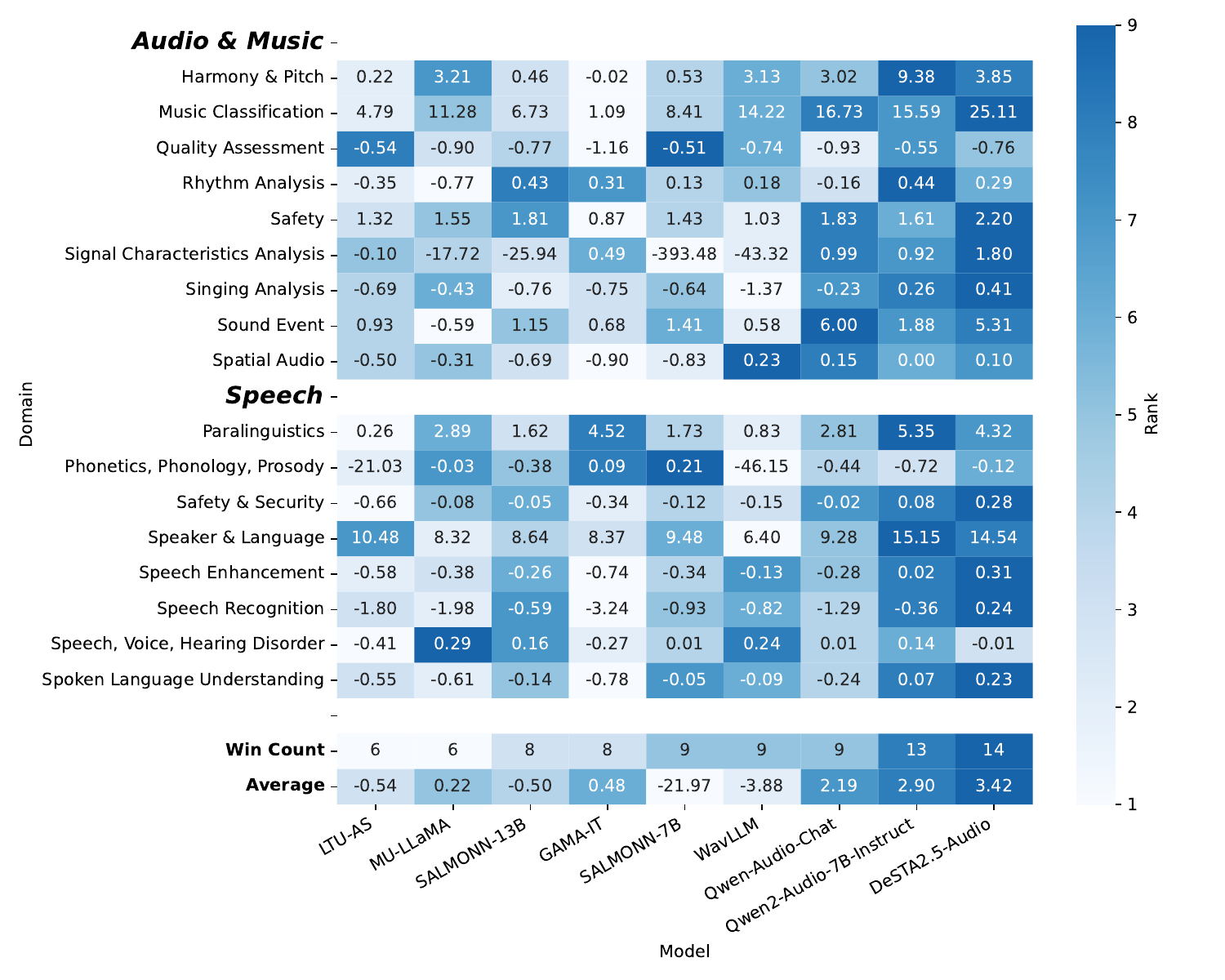}
\caption{LALM performance on the Dynamic-SUPERB Phase-2 benchmark \cite{huang2025dynamicsuperb}, relative to the ASR+LLM baseline. Each cell shows a model’s relative score in a specific domain, with darker blue indicating a higher performance rank. The bottom rows summarize each model’s domain win count (i.e., positive scores) and average performance across all domains.}

\label{fig:dynamic-superb2}
\end{figure*}

\subsubsection{\textbf{Overall Trend}}
\hfill

%

The results reveal a consistent trend in model rankings across multiple benchmarks. Notably, DeSTA2.5-Audio demonstrates consistently superior performance, emerging as the best-performing model with strong generalization across diverse audio-language tasks. It achieves the highest scores on Dynamic-SUPERB Phase-1 (69.53), MMAU (57.50), SAKURA-Multi (69.85), and Speech-IFEval (93.89), indicating its robustness and generalization across multiple domains and conditions.

Compared to DeSTA2, which focused primarily on speech data, DeSTA2.5-Audio shows improved performance attributed to the broader coverage of environmental sounds and musical audios in addition to speech. This demonstrates that our proposed self-generated data construction strategy is general and can be effectively extended to other domains.
Interestingly, while Qwen2-Audio-Instruct leveraged a massive 510,000 hours of training data, DeSTA2.5-Audio achieves competitive or superior performance using only 7,000 hours. This highlights the efficiency and effectiveness of our training methodology, which enables strong results with significantly less data.

The cascade baselines (ASR+LLM and ASR+AAC+LLM), while limited to text-based inputs such as transcriptions or audio captions, still perform surprisingly well.
In some benchmarks, they even outperform earlier LALMs, such as LTU-AS, SALMONN, and Qwen-Audio. 
This suggests that an LLM can effectively interpret audio concepts when represented in textual form.
Although some fine-grained acoustic information is inevitably lost, these baselines show strong instruction-following ability.
In contrast, DeSTA2.5-Audio adopts continuous representations of audio, enabling the model to capture richer and more nuanced information beyond what text alone could convey.
During the alignment process, the fusion model learns to interact directly with these audio representations, resulting in more grounded and fine-grained perception.
This approach allows DeSTA2.5-Audio to outperform cascade baselines, especially on tasks that require deeper audio understanding.

\subsubsection{\textbf{Task Performance}}
\hfill

DeSTA2.5-Audio consistently achieves the best performance across a variety of subdomains. Within speech-related domains, the model demonstrates strong capabilities not only in content and semantic comprehension but also in more advanced perception, such as paralinguistics, degraded speech, and speaker identity in the Dynamic-SUPERB phase-1. Beyond speech, DeSTA2.5-Audio also performs competitively in the audio and music tracks of the MMAU benchmark, underscoring its versatility in processing diverse auditory inputs.

In the single-hop setting (SAKURA-Single), which requires grounding questions to specific audio segments, Qwen2-Audio-Instruct outperforms DeSTA2.5-Audio (81.20 vs. 76.65). This is likely attributable to Qwen2-Audio-Instruct’s extensive instruction tuning, which enhances its ability to respond to direct queries.
However, the performance dynamics shift dramatically in the multi-hop setting (SAKURA-Multi). Here, DeSTA2.5-Audio emerges as the top performer with a score of 69.85, while other models suffer significant drops. For example, Qwen2-Audio-Instruct declines by over 30 points (from 81.20 to 49.10), and LTU-AS drops from 40.90 to 18.10, underscoring their struggles with compositional reasoning.
A similar trend can be observed in the Speech-IFEval benchmark. For instance, models such as LTU-AS and SALMONN exhibit low IFrate and high forgetting rates, indicating a significant degradation in both text-based knowledge and instruction-following capabilities. 
In contrast, DeSTA2.5-Audio and DeSTA2 demonstrate remarkable robustness on the Speech-IFEval benchmark. Their performance remains comparable to or even surpasses the text-based reference systems~\cite{lu2025speechifeval}. These benchmarks demand extensive textual knowledge and complex instruction-following, effectively serving as out-of-domain tests for the LALMs.
The results show that our proposed approach yields a significant advantage in multi-hop reasoning and instruction-following performance, while also delivering strong results on core auditory perception tasks.

\begin{table*}[!t]
\renewcommand{\arraystretch}{1.3}
\caption{Evaluation results on VoiceBench. Performance is measured using task-specific metrics, where higher scores indicate better performance.}
\label{tab:voicebench_results}
\setlength{\tabcolsep}{4pt}
\centering
\begin{tabular}{lcccccccccc}
\toprule
\textbf{Model} & \textbf{AlpacaEval} & \textbf{CommonEval} & \textbf{WildVoice} & \textbf{SD-QA} & \textbf{MMSU}
& \textbf{OpenbookQA} & \textbf{BBH} & \textbf{IFEval} & \textbf{AdvBench} & \colorbox{lightgray}{\textbf{Overall}} \\

\midrule
ASR+LLM & 4.53 & 4.04 & 4.16 & 70.43 & 62.43 & 72.53 & 69.70 & 69.53 & 98.08 & 77.48 \\
GPT-4o-Audio~\cite{achiam2023gpt}
& 4.78 & 4.49 & 4.58 & 75.50 & 80.25 & 89.23 & 84.10 & 76.02 & 98.65 & 86.75 \\
\midrule
Moshi \cite{defossez2024moshi} & 2.01 & 1.60 & 1.30 & 15.64 & 24.04 & 25.93 & 47.40 & 10.12 & 44.23 & 29.51 \\

GLM-4-Voice \cite{zeng2024glm} & 3.97 & 3.42 & 3.18 & 36.98 & 39.75 & 53.41 & 52.80 & 25.92 & 88.08 & 56.48 \\

DiVA \cite{held2024distilling}
& 3.67 & 3.54 & \underline{3.74} & \underline{57.05} & 25.76 & 25.49 & 51.80 & 39.15 & \underline{98.27} & 57.39 \\

Qwen2-Audio-Instruct \cite{chu2024qwen2} & 3.74 & 3.43 & 3.01 & 35.71 & 35.72 & 49.45 & 54.7 & 26.33 & 96.73 & 55.80 \\

Phi-4 \cite{abouelenin2025phi} & 3.81 & \underline{3.82} & 3.56 & 39.78 & 42.19 & 65.93 & 61.8 & 45.35 & \textbf{100.00} & 64.32 \\

VITA-1.5 \cite{fu2025vita} & 4.21 & 3.66 & 3.48 & 38.88 & 52.15 & \underline{71.65} & 55.30 & 38.14 & 97.69 & 64.53 \\

DeSTA2 \cite{desta2} & \underline{4.36} & 3.33 & 3.42 & 56.06 & \textbf{66.43} & 61.98 & \underline{66.2} & \underline{52.63} & \underline{98.27} & \underline{69.31} \\

DeSTA2.5-Audio
& \textbf{4.41} & \textbf{3.99} & \textbf{3.80} & \textbf{60.04} & \underline{60.87} & \textbf{74.06} & \textbf{67.00} & \textbf{66.42} & \underline{98.27} & \textbf{74.52}

\\
\bottomrule
\end{tabular}
\end{table*}

\subsection{Results on Dynamic-SUPERB Phase-2}
\label{sec:ds2_results}

Figure~\ref{fig:dynamic-superb2} presents results on the Dynamic-SUPERB Phase-2 benchmark, which comprises 180 diverse, crowd-sourced tasks, each evaluated using task-specific metrics. To support intuitive visualization and comparison, the benchmark organizes tasks into domains and reports relative scores with respect to an ASR+LLM cascade baseline, which serves as the 0\% reference point. This framing emphasizes model capabilities beyond those of a basic cascade pipeline.

Compared to other LALMs, DeSTA2.5-Audio achieves the highest overall performance, ranking first in both win count (14 domains) and average relative score (3.42). These results highlight the model’s robustness and versatility across a wide range of general audio tasks. Notably, DeSTA2.5-Audio does not rely on task-specific instruction tuning. Instead, we leverage diverse and descriptive audio–text pairs, allowing the model to develop a multifaceted understanding of audio while keeping its language ability.

A closer analysis reveals that in certain domains, such as ``Quality Assessment, Speech, Voice, Hearing Disorder", and ``Phonetics, Phonology, and Prosody", our model underperforms the cascade baseline, as indicated by negative relative scores. 
This suggests that the audio attributes required for these tasks are currently underrepresented in our training corpus, thereby limiting generalization performance in these areas.
However, the relative scores remain relative to the ASR+LLM baseline, implying marginal deviation despite underperformance. In contrast, some systems exhibit severe performance degradation, for example, LTU-AS in Phonetics, Phonology, and Prosody, and SALMONN in Signal Characteristics Analysis.

Interestingly, rather than producing hallucinated or misleading responses in unfamiliar tasks, DeSTA2.5-Audio often outputs responses like ``I don't have enough information," demonstrating a form of uncertainty awareness. We consider this behavior particularly valuable for real-world deployment scenarios, where the ability to recognize and communicate uncertainty is critical for trustworthiness and reliability. This trait can be attributed to both the inherent behavior of the backbone LLM and our training strategy, which implicitly encourages such behavior. 

\subsection{Results on VoiceBench}
\label{sec:voice_results}

Table~\ref{tab:voicebench_results} presents the results on VoiceBench, where the audio input consists of spoken instructions, such as questions or commands, and the model is expected to generate appropriate responses. Unlike previous benchmarks that focus mainly on auditory understanding or instruction-following, VoiceBench emphasizes general knowledge and interactive behavior. 
To align with these goals, we guide DeSTA2.5-Audio using a system prompt that encourages more natural and conversational responses.

In terms of overall performance, the ASR+LLM pipeline proves to be a very strong baseline, achieving an overall score of 77.48 and outperforming most fusion-based approaches. This indicates that the tasks in VoiceBench are primarily content-driven and can be effectively addressed by accurate transcription paired with a powerful LLM. For comparison, OpenAI’s proprietary GPT-4o-Audio achieves the highest score of 86.75 on VoiceBench, establishing the current state-of-the-art in the field.
Despite the simplicity of cascade systems, most open-source fusion models are still unable to exceed this baseline. This highlights a notable research gap in the field of recent LALM development. These models differ considerably in how they represent audio input. For example, Moshi~\cite{defossez2024moshi} and GLM-4-Voice~\cite{zeng2024glm} adopt unit-based representations, while DiVA~\cite{held2024distilling}, Qwen2-Audio-Instruct~\cite{chu2024qwen2}, Phi-4~\cite{abouelenin2025phi}, VITA-1.5~\cite{fu2025vita} use continuous representations. Furthermore, several models such as Moshi, GLM-4-Voice, and VITA-1.5 are specifically designed for speech-to-speech interaction, making their training procedure more difficult than other speech-to-text models.
Among all fusion-based models evaluated, DeSTA2.5-Audio achieves an overall score of 74.52, outperforming models such as VITA-1.5 (64.53) and Qwen2-Audio-Instruct (55.80). It demonstrates balanced and consistent results across a broad range of subtasks, including AlpacaEval, CommonEval, and OpenbookQA. 
This performance is largely attributed to the strong language ability of DeSTA2.5-Audio, where it not only follows the system prompt to adapt its behavior into an interactive style, but also responds appropriately based on its inherent knowledge.

In summary, the proposed DeSTA2.5-Audio model demonstrates state-of-the-art performance or comparable across a wide range of audio-language benchmarks. It outperforms both previous versions (\textit{e.g.}, DeSTA2) and competing models on key tasks such as Dynamic-SUPERB Phase-1, MMAU, SAKURA-Multi, and Speech-IFEval (Section~\ref{sec:big_results}). Notably, DeSTA2.5-Audio achieves competitive results with significantly less training data (7,000 hours) compared to models like Qwen2-Audio-Instruct (510,000 hours). It also exhibits robustness in knowledge-intensive tasks and shows strong instruction-following ability. Evaluations on Dynamic-SUPERB Phase-2 (Section~\ref{sec:ds2_results}) and VoiceBench (Section~\ref{sec:voice_results}) further validate its generalization and robustness, highlighting its potential as a general-purpose LALM.

\subsection{Representation-Level Analysis}
\label{sec:representation_level_analysis}
\begin{table}[]
    \centering
    \color{blue}
    \caption{
    \color{blue}
    Comparison of instruction-based QA and similarity-based classification on SAKURA-Single. Higher value indicates better cross-modal alignment quality.}
    \begin{tabular}{lccc}
    \toprule
         Method &  Gender & Emotion & Animal \\
         \# classes & 2 & 5 & 9 \\
         \midrule
         Random & 50.0 & 20.0 & 11.1 \\
         Instruction-based QA & 89.2 & \textbf{62.2} & 58.8 \\
         Similarity-based classification & \textbf{95.2} & 56.2 & \textbf{66.2} \\
    \bottomrule 
    \end{tabular}
    
    \label{tab:zero-shot-classification}
\end{table}


{\color{blue}

Beyond standard instruction-following evaluations, we investigate the cross-modal alignment quality between audio and text representations within DeSTA2.5-Audio, as described in Section~\ref{sec:analysis_study}. Table~\ref{tab:zero-shot-classification} compares standard text generation (``Instruction-based QA'') with implicit representation alignment (``Similarity-based classification''). The results demonstrate that DeSTA2.5-Audio establishes robust cross-modal grounding and encodes highly discriminative audio features, showing performance comparable to instruction-based evaluation. This indicates that DeSTA2.5-Audio successfully integrates audio representations into the backbone LLM's semantic space. Furthermore, our results reveal an interesting trade-off. For purely acoustic tasks like Gender and Animal sounds, the model's internal representations outperform text generation. This suggests that some audio information may be slightly diluted or lost during the instruction-following and text decoding stages. However, the opposite occurs for Emotion. Since emotion depends not just on acoustic features but also on spoken content, the representation-based approach may miss semantic nuances, whereas the instruction-based approach can leverage them for more accurate emotion recognition.

\begin{figure}
    \centering
    \includegraphics[width=0.98\linewidth]{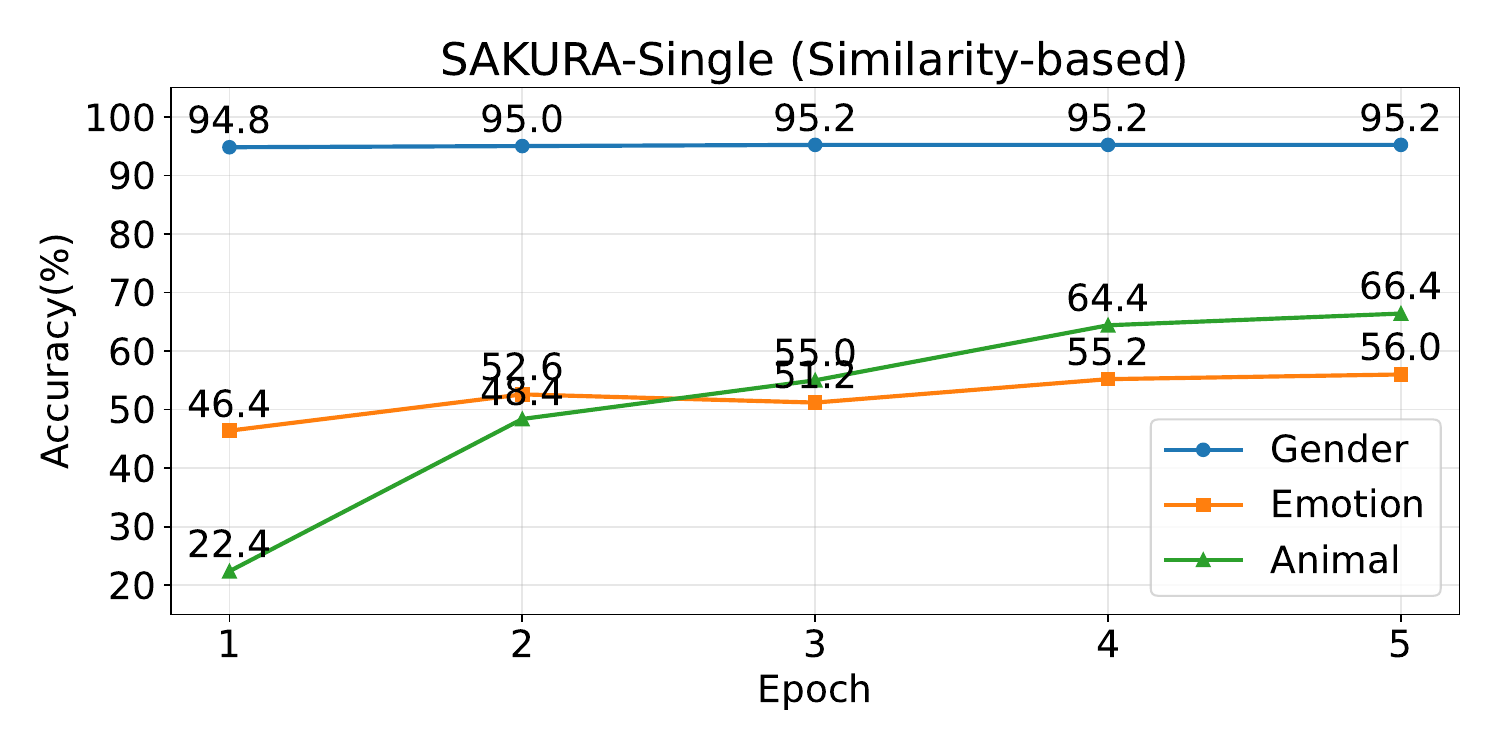}
    \caption{
    \color{blue}
    Epoch-wise evolution of cross-modal alignment. The curves depict similarity-based classification accuracy during training.
    }
    \label{fig:accuracy_per_epoch}
\end{figure}

\begin{figure}
    \centering
    \includegraphics[width=0.94\linewidth]{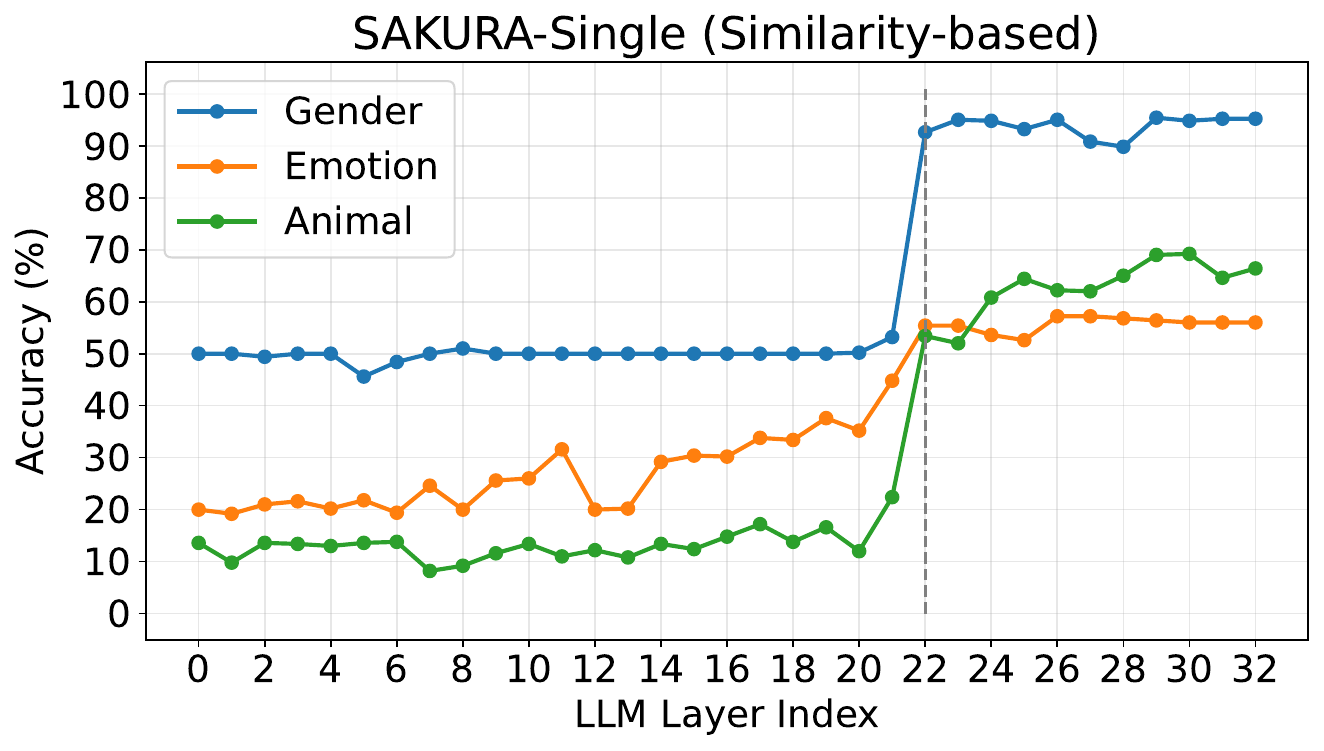}
    \caption{
    \color{blue}
    Layer-wise analysis of cross-modal alignment within the LLM. We measure similarity-based classification accuracy using hidden states from each transformer layer. The sharp rise at the deeper layers indicates that audio-text integration occurs through a deferred process within the semantic space.}
    \label{fig:layer_wise_accuracy}
\end{figure}

Figure~\ref{fig:accuracy_per_epoch} illustrates the evolution of alignment quality across training epochs. We observe that simple acoustic attributes, such as gender, are learned rapidly, achieving near-convergence by the first epoch and more complex concepts like emotion and animal sounds demonstrate a steady upward trend as training progresses. This trajectory confirms that DeSTA2.5-Audio effectively drives the model to progressively refine its understanding of multifaceted auditory concepts over time.

Figure~\ref{fig:layer_wise_accuracy} presents a detailed analysis of the accuracy based on hidden states extracted from different layers of the backbone LLM. Interestingly, a distinct phase transition phenomenon is observed. Accuracy remains near random levels in the shallow layers, but exhibits a sharp surge around layer 22. 
This finding implies that the model does not enforce immediate audio-text alignment in the input layers. Instead, the end-to-end optimization allows the raw acoustic features to be progressively integrated into the LLM’s semantic space in the deeper layers. 
This pattern is consistently observed in the comparative experiments in Section~\ref{sec:comparison_studies} as well.
We hope these findings inspire broader investigation into the underlying dynamics of cross-modal learning, fostering a deeper understanding of interpretability in multimodal foundation models.

}

\begin{table*}[!t]
\renewcommand{\arraystretch}{1.3}
\setlength{\tabcolsep}{2.5pt}
\caption{Evaluation results across various configurations on representative benchmarks. ``Backbone" and ``Dataset" refer to the LLM used for LALM training and for dataset generation, respectively. The label ``(1-P)" indicates that the dataset was generated using a single descriptive prompt\cite{desta2}. ``PPL" denotes the perplexity of the training targets as computed by the corresponding backbone LLM. ``N/A" indicates that the result is unavailable due to degeneration.}
\label{tab:comparison_study}
\centering
\begin{tabular}{lllccccccccccccccc}
\toprule
\multirow{2}{*}{\textbf{ID}} & \multirow{2}{*}{\textbf{Backbone}} & \multirow{2}{*}{\textbf{Dataset}} & \multirow{2}{*}{\textbf{PPL}} & \multicolumn{6}{c}{\textbf{Dynamic-SUPERB Phase-1}} 
& \multicolumn{4}{c}{\textbf{MMAU-test-mini}} 
& \multicolumn{2}{c}{\textbf{SAKURA}}
& \multicolumn{2}{c}{\textbf{Speech-IFEval}} \\

\cmidrule(lr){5-10}
\cmidrule(lr){11-14}
\cmidrule(lr){15-16}
\cmidrule(lr){17-18}
 & & & & CON & SEM & PAR & DEG & SPK & \colorbox{lightgray}{Avg} 
 & Speech & Sound & Music & \colorbox{lightgray}{Avg} 
 & Single & Multi 
 & IFrate & $\Delta$ \\
\midrule
\multicolumn{2}{l}{\textbf{Dataset configuration}} \\
\midrule

\multicolumn{2}{l}{\textit{Self-generation}} \\

A1 & Llama3.1 & Llama3.1
& 1.61
& 89.91 & 70.50 & 59.50 & 56.89 & 52.70 & 66.10
& 57.36 & 57.66 & 48.20 & 54.40
& 72.45 & 61.25
& \textbf{91.51} & -2.15
\\

A2 & Qwen2.5 & Qwen2.5
& 1.27
& 87.45 & 65.58 & 46.86 & 72.42 & 59.50 & \textbf{69.94}
& 54.05 & 65.77 & 52.10 & \textbf{57.30}
& \textbf{78.30} & \textbf{68.65}
& 88.49 & \textbf{-0.24}
\\

A3 & Llama3.1 & Llama3.1 (1-P)
& 1.58
& 79.73 & 60.33 & 54.57 & 55.95 & 55.60 & 61.71
& 51.95 & 56.76 & 41.62 & 50.10
& 55.95 & 49.85
& 90.09 & -3.67 
\\

\midrule
\multicolumn{2}{l}{\textit{External-model}} \\

B1 & Llama3.1 & Qwen2.5
& 3.54
& N/A & N/A & N/A & N/A & N/A & N/A 
& N/A & N/A & N/A & N/A 
& N/A & N/A 
& N/A & N/A 

\\

B2 & Llama3.1 & Gemma3-12B
& 4.85
& 72.55 & 70.33 & 39.86 & 54.26 & 52.50 & 58.18
& 55.86 & 48.65 & 38.02 & 47.50
& 52.70 & 32.25
& \textbf{90.92} & \textbf{-2.79}
\\

B3 & Llama3.1 & Llama3.1-70B
& 2.20
& 93.77 & 73.25 & 54.64 & 62.74 & 57.50 & \textbf{69.44}
& 54.95 & 56.46 & 44.61 & \textbf{52.00}
& \textbf{75.70} & \textbf{67.35}
& 89.73 & -4.06

\\

\midrule
\multicolumn{2}{l}{\textbf{Model configuration}} \\
\midrule
\multicolumn{2}{l}{\textit{With LoRA Adapter}} \\
C1 & Llama3.1-LoRA & Llama3.1
& 1.61
& 83.00 & 66.00 & 57.50 & 61.82 & 61.40 & 66.52
& 60.06 & 53.45 & 52.99 & 55.50
& \textbf{71.50} & \textbf{64.85}
& \textbf{91.17} & \textbf{-2.52}
\\

C2 & Llama3.1-LoRA & Qwen2.5
& 3.54
& 87.36 & 72.92 & 45.43 & 70.82 & 54.10 & \textbf{69.43}
& 59.16 & 61.26 & 46.41 & \textbf{55.60}
& 58.75 & 52.85
& 84.07 & -10.11
\\





\bottomrule
\end{tabular}
\end{table*}

\section{Comparison Studies}
\label{sec:comparison_studies}
\begin{figure}
    \centering
    \includegraphics[width=0.98\linewidth]{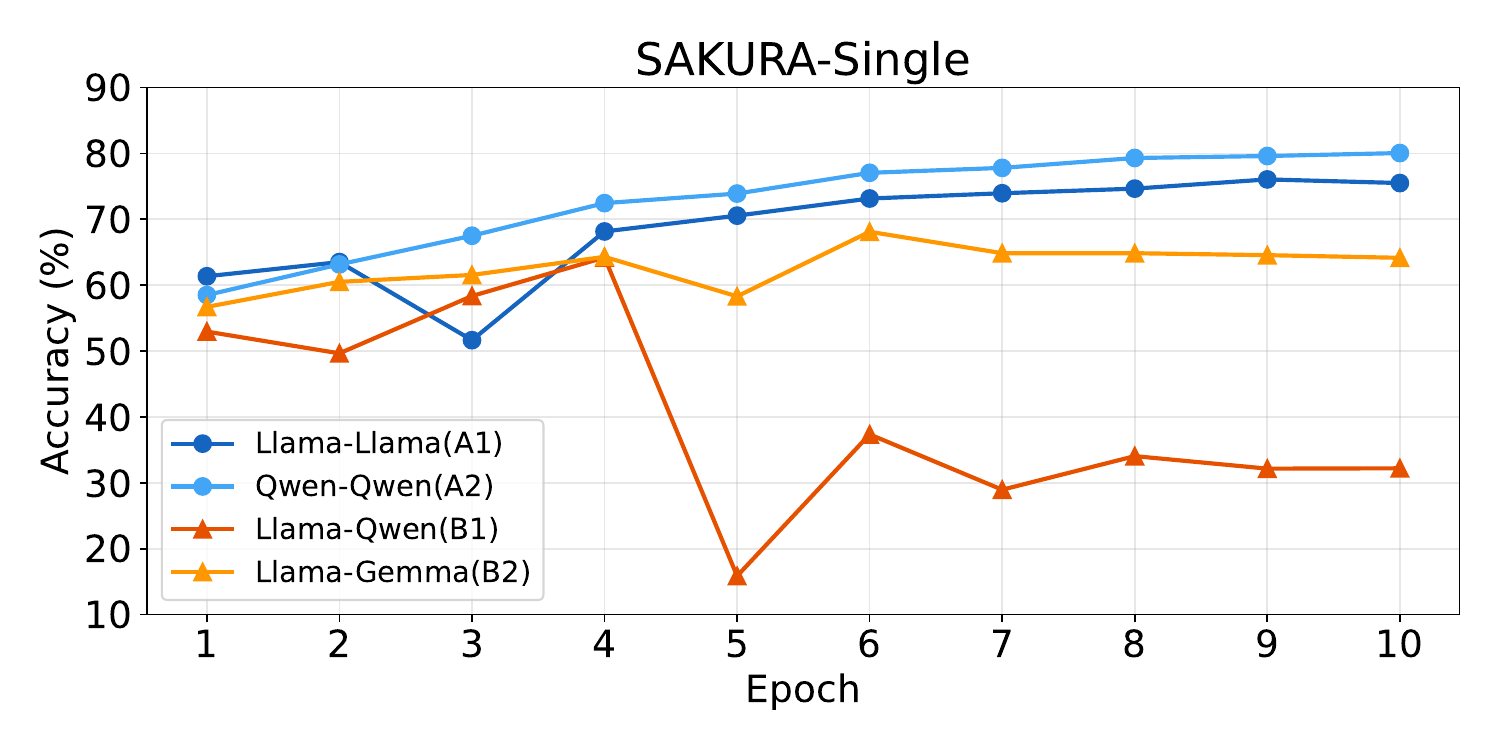}
    \caption{
    \color{blue}
    Epoch-wise evolution of different data construction settings on instruction-based SAKURA-Single, which requires combined audio perception and instruction-following abilities. } 
    \label{fig:training_dynamics}
    \vspace{-1em}
\end{figure}

{\color{blue}
As shown in Table \ref{tab:comparison_study}, we evaluate the  self-generation approach under controlled experimental setups using identical audio-description pairs from a 500K subset of DeSTA-AQA5M. We compare the self-generation setting (backbone-generated targets) against external-model settings using targets from other LLMs. Notably, the external-model setup simulates a distribution mismatch, such as when human annotators or alternative LLMs are employed to create the training corpus, which is widely adopted in previous LALM studies~\cite{lu24c_interspeech, gong2023joint, gonglisten, tang2024salmonn, ghosh-etal-2024-gama, hu2024wavllm, pmlr-v235-kong24a, chu2023qwen,chu2024qwen2, abouelenin2025phi, huang2025speechcaps, ghosh2025audio}. We also analyze the performance evolution in Figure~\ref{fig:training_dynamics} and cross-modal alignment quality in Table~\ref{tab:ablation_zero-shot-classification}. Building upon these two setups, we explore configurations with LoRA adapters integrated into the backbone LLM. 
To quantify the degree of distribution mismatch, we report the perplexity (PPL) of the training targets as computed by the corresponding backbone LLM. A lower perplexity indicates the model is more likely to produce the training target sequence, suggesting less distributional discrepancy in the dataset.
}

\subsection{Comparison between Self-generation and External-model}
\label{sec:self-output}

{
\color{blue}

As indicated in Table~\ref{tab:comparison_study} and Figure~\ref{fig:training_dynamics}, the self-generated setup exhibits consistently superior performance and training stability. Since the training targets are consistent with the backbone's native output distribution, the optimization process avoids conflicting stylistic adaptation, focusing exclusively on cross-modal alignment. 
When comparing Llama3.1 (A1) and Qwen2.5 (A2), Qwen2.5 consistently outperforms Llama3.1 across all benchmarks. This performance gap may be attributed to Qwen2.5’s stronger text generation capabilities. 
While Qwen2.5’s performance in basic content comprehension tasks is relatively on par with Llama3.1, it demonstrates better performance in other areas, such as speaker category in Dynamic-SUPERB Phase-1, and environmental sounds and music understanding in MMAU.
Previous evaluations on text-based benchmarks have also shown that Qwen2.5 exhibits superior reasoning and mathematical abilities compared to Llama3.1~\cite{open-llm-leaderboard-v2}. 
However, there is currently no conclusive evidence indicating a corresponding advantage in auditory perception, which warrants further investigation. Nevertheless, under identical training conditions, our experimental results suggest that Qwen2.5 serves as a more effective backbone LLM than Llama3.1. These findings also indicate that our training framework generalizes well across different LLMs.
}

{\color{blue}
In contrast, external-model settings introduce distributional shifts that severely disrupt training. Figure~\ref{fig:training_dynamics} illustrates the severe consequences of this conflict. In the B1 setting (Llama3.1 trained on Qwen2.5-generated data), we observe catastrophic model degeneration~\cite{shumailov2024ai} after the fourth epoch, manifesting as repetitive or nonsensical outputs. Despite Qwen2.5-generated data proving effective in self-generated experiments (A2), this failure suggests that the discrepancy destabilizes the training process. Similarly, while the B2 setting (Llama3.1 trained on Gemma3-generated data) avoids model collapse, the learning curve plateaus prematurely, indicating that the distributional discrepancy still hinders the model's ability to refine auditory perception.

\begin{table}[]
    \centering
    \color{blue}
    \caption{
    \color{blue}
    Alignment quality analysis on SAKURA-Single comparing instruction-based QA (Inst.) and similarity-based classification (Sim.). 
L, Q and G denote Llama, Qwen and Gemma, respectively.}
    \begin{tabular}{l|cc|cc|cc}
    \toprule
         \textbf{ID} &  \multicolumn{2}{c|}{\textbf{Gender}} & \multicolumn{2}{c|}{\textbf{Emotion}} & \multicolumn{2}{c}{\textbf{Animal}} \\
          & \textbf{Inst.} & \textbf{Sim.} & \textbf{Inst.} & \textbf{Sim.} & \textbf{Inst.} & \textbf{Sim.} \\
    \midrule
         Random & \multicolumn{2}{c|}{50.0} & \multicolumn{2}{c|}{20.0} & \multicolumn{2}{c}{11.1} \\
         \midrule
         A1 (L-L)  & \textbf{91.0} & \textbf{93.0} & 57.2 & \textbf{55.6} & 57.4 & 65.4 \\
         A2 (Q-Q)  & 89.8 & 86.8 & \textbf{62.2} & 54.8 & \textbf{77.8} & \textbf{77.0} \\
         B1 (L-Q)  & 28.0 & 58.6  & 5.8 & 20.4 & 30.4 & 15.2 \\
         B2 (L-G)  & 62.4 & 83.4  & 42.0 & 34.4 & 24.2 & 26.2 \\
         
    \bottomrule 
    \end{tabular}
    
    \label{tab:ablation_zero-shot-classification}
\end{table}

To determine whether these failures arise from poor alignment or generation issues, Table~\ref{tab:ablation_zero-shot-classification} presents an analysis based on instruction-following and similarity-based metrics. The self-generated settings (A1, A2) achieve high scores in both metrics, demonstrating successful cross-modal alignment that maps audio features to semantic concepts within the representation space. Consequently, the model effectively leverages its intrinsic instruction-following capabilities to jointly process audio inputs and text instructions, facilitating robust instruction-based evaluation.
Conversely, the external-model settings (B1, B2) exhibit significant degradation in both areas. Given that the training target is the sole variable, these results reveal that distributional discrepancy not only undermines training stability but also impedes the effective integration of audio information, directly leading to the poor downstream performance observed across all metrics in Table~\ref{tab:comparison_study}.

Finally, comparing A1 and A3 (Table ~\ref{tab:comparison_study}) highlights the benefits of prompt diversity within the self-generated framework. Utilizing diverse prompts (A1) rather than a single fixed prompt (A3) enriches data diversity, thereby enhancing generalization. Notably, these strong zero-shot generalization results are achieved without requiring any task-specific instruction pairs. Even when the training data is constructed using only randomly sampled prompts, the model successfully leverages the inherent capabilities of the LLM backbone to perform well.
Furthermore, we explore the impact of model strength by using Llama3.1-70B to generate training targets (B3). Although the lower perplexity (2.20) suggests a closer data distribution, this setting does not consistently yield improvements across all tasks compared to the self-generated experiment (A1).
}

{
\color{blue}
\subsection{With LoRA Adapter}
In the LoRA adapter setting, we introduce trainable parameters to the backbone LLM, which is expected to increase model capacity and help mitigate the distribution mismatch problem. In the self-generation setup (C1), we find that adding LoRA layers yields similar or only slightly better performance. This indicates that, under self-generation settings, incorporating LoRA adapters does not provide significant advantages. In other words, fine-tuning a lightweight modality adapter is sufficient for cross-modal alignment.
Interestingly, when training with Qwen2.5-generated data (C2), performances on audio processing tasks, such as Dynamic-SUPERB, MMAU, and SAKURA-Single, is comparable to the self-generation setup (A2). However, it experiences a significant degradation in SAKURA-Multi and Speech-IFEval, which requires additional text knowledge and instruction-following ability.

This disparity indicates that while LoRA adapters can mitigate distribution mismatch for in-domain tasks, they fail to preserve broader capabilities on out-of-domain benchmarks, suggesting that the primary utility of LoRA in such scenarios is to encode distributional mismatch between external targets and the backbone's native behavior.
This finding may provide a rationale for why previous works  observed that adjusting or pruning LoRA weights during inference can help recover instruction-following ability~\cite{tang2024salmonn,lu24c_interspeech}.
While introducing LoRA adapters remains a common approach in recent LALM development, our results suggest that reducing the discrepancy between training data and the model’s native distribution is a more critical factor for preserving generalization than architectural modifications alone.
}

\section{Conclusion}
{\color{blue}
In this work, we introduced DeSTA2.5-Audio, a general-purpose LALM built on the self-generated cross-modal alignment framework, DeSTA. By enabling the backbone LLM to generate its own training targets, we eliminate distributional shifts in the training dataset, thereby mitigating the catastrophic forgetting problem commonly observed in prior works. Under this framework, we successfully develop a robust and generalizable LALM without relying on task-specific instruction data.
To support this, we constructed DeSTA-AQA5M, a large-scale, task-agnostic dataset comprising 5 million audio-text pairs from 50 diverse audio datasets. Trained solely on this corpus, DeSTA2.5-Audio achieves state-of-the-art or competitive results across benchmarks such as Dynamic-SUPERB (Phase-1 and Phase-2), MMAU, SAKURA, Speech-IFEval, and VoiceBench. Our analytical study reveals that the model effectively maps audio information into semantic concepts within the LLM representation space. This successful alignment enables the model to leverage its intrinsic instruction-following capabilities, thereby facilitating robust performance on downstream instruction-based benchmarks.
Our findings introduce a new training philosophy for future LALM development, highlighting the importance of training data quality and consistency over quantity. Compared to previous studies, our self-generation strategy offers a scalable and robust solution with superior generalization.

While our current design leverages text descriptions as an intermediate bridge, we acknowledge that not all acoustic nuances can be effectively captured through textual representations. In future work, we will explore audio-to-audio supervision and preference learning from human feedback to better capture subtle, non-text-expressible acoustic features, advancing toward LALMs with more complete and human-like audio perception.
}

\section{Acknowledgment}
The authors thank NVIDIA Taiwan AI R\&D Center for the TRDC budget support and contributions to this research.
This work has also been supported by NVIDIA Academic Grant from the NVIDIA Developer Program 2025.

\bibliographystyle{IEEEtran}
\bibliography{ref}

\appendix

\subsection{Author Contributions}

All authors contributed meaningfully to the design of the method, data collection, evaluation, writing, and refinement of the paper. While all authors were involved in multiple aspects of the project, their primary contributions are highlighted below:

\textbf{Ke-Han Lu} proposed the initial idea and led the overall project direction. He conducted the majority of the experiments and was primarily responsible for drafting the manuscript.
\textbf{Zhehuai Chen, Szu-Wei Fu, Chao-Han Huck Yang, Sung-Feng Huang, Boris Ginsburg, Yu-Chiang Frank Wang, and Hung-yi Lee} contributed deep technical expertise, helped shape the research direction, and provided crucial guidance and feedback on experimental methodology and manuscript development. Furthermore, their contributions to our earlier works~\cite{lu24c_interspeech, desta2} were served as a critical stepping stone for the conception and execution of this study.
\textbf{Kai-Wei Chang and Cheng-Han Chiang} offered valuable discussion during the early stages of the project and played a significant role in refining the experiments and manuscript.
\textbf{Chih-Kai Yang, Chee-En Yu, Chun-Wei Chen, Wei-Chih Chen, and Chien-yu Huang} conducted the majority of evaluations for both the proposed model and baseline systems.
\textbf{Chih-Kai Yang, Yi-Cheng Lin, Yu-Xiang Lin, Chi-An Fu, Chun-Yi Kuan, Wenze Ren, Xuanjun Chen, Wei-Ping Huang, En-Pei Hu, Tzu-Quan Lin, Yuan-Kuei Wu, Kuan-Po Huang, Hsiao-Ying Huang, and Huang-Cheng Chou} contributed significantly to the data collection effort, which was critical to the success of this project. They gathered, curated, and consolidated diverse datasets from multiple research domains into a unified and structured format, laying a solid foundation for effective experimentation and thorough analysis.

\newpage

\end{document}